\documentclass[a4paper,twocolumn,10pt]{article}
\usepackage[english]{babel}
\usepackage[latin1]{inputenc}
\usepackage[T1]{fontenc}
\usepackage{a4wide,latexsym,verbatim}
\usepackage{amsmath,amssymb,amsfonts}
\usepackage{color}
\usepackage[table]{xcolor}
\definecolor{gr}{rgb}{0.7,0.7,0.7}
\definecolor{ulb}{rgb}{0,0.27,0.55}
\definecolor{new}{rgb}{0.21,0.43,0.5}

\usepackage[pdfauthor={Michael V. Bebronne},pdftitle={Quasi-linear static solutions in massive gravity},pdftex]{hyperref}
\hypersetup{colorlinks=true,urlcolor=ulb,linkcolor=ulb,citecolor=ulb}
\usepackage{graphpap,graphicx}
\newcommand{\dif}{\textrm{d}}
\newcommand{\tx}[1]{\textrm{#1}}
\newcommand{\Mpl}{\textrm{M}_{\textrm{pl}}^2}
\newcommand{\Mplu}{\textrm{M}_{\textrm{pl}}}

\usepackage[text={19cm,25cm},centering]{geometry}
\usepackage{chngpage}

\begin{document}

\twocolumn[ \begin{@twocolumnfalse}
\vspace{0cm}

{\fontsize{20}{80}\usefont{OT1}{pzc}{m}{n}\selectfont Quasi-linear static solutions in massive gravity}
\vspace{1cm} \\
Michael V. Bebronne \vspace{0.1cm}\\
{\fontsize{8}{80} \textit{Service de Physique Th\'eorique, Universit\'e Libre de Bruxelles (ULB), \\ CP225, Boulevard du Triomphe, 1050 Brussels, Belgium.}}

\vspace{0.5cm}

\begin{tabular}{ll}
{\color{gr}{\rule{4.1cm}{0.03cm}}} & {\color{gr}{\rule{14cm}{0.03cm}}} \\
\begin{minipage}{4.1cm}
\vspace{1.3cm} 
{\fontsize{8}{80}\texttt{ULB reference:}\\
\textit{ULB-TH/10-17}
\vspace{0.1cm} \\
\texttt{e-mail address:}\\
\textit{mbebronn@ulb.ac.be}
\vspace{0.1cm} \\
\texttt{Published in:} \\
\textit{Physical Review D 82, 024020}}
\end{minipage} &
\begin{minipage}{14cm}
\vspace{0.3cm}
\textsc{A b s t r a c t} \vspace{0.1cm} \\
{\small The static vacuum spherically symmetric solutions of massive gravity theories possess two integration constant: the mass M and a scalar charge S. The presence of this scalar charge reflects the modification of the gravitational interaction as compared to General Relativity. Surprisingly, these solutions have an asymptotic behavior different from that obtained in the linear perturbation theory. The aim of this paper is to understand how these modified spherically symmetric solutions emerge from a quasi-linear approximation in order to generalize them to any arbitrary mass distribution. Along with these modified solutions, we found a new class of static solutions having a Yukawa shape.}
\end{minipage} \\
{\color{gr}{\rule{4.1cm}{0.03cm}}} & {\color{gr}{\rule{14cm}{0.03cm}}}
\end{tabular}

\vspace{1cm}

\end{@twocolumnfalse}]

\section{Introduction}
\label{sc:intro}

Gravity is the most familiar but also the most mysterious of the four fundamental interactions. Described by Einstein's theory of General Relativity (GR), gravity has been tested in the weak field limit up to post-Newtonian corrections \cite{Will:2005va}. Although indirect detection of gravitational waves seems to be in agreement with GR \cite{1979Natur.277..437T,Taylor:1982zz}, recent advances in observational cosmology \cite{Perlmutter:1998np,Dunkley:2008ie,AdelmanMcCarthy:2007wu} have revived interest in alternative theories of gravity in which the gravitational interaction is modified at very large distances. The motivations behind these theories lie in the possibility of explaining the observations without introducing the dark energy and matter components. Different approaches to this problem have been discussed in the literature \cite{Milgrom:1983pn,Bekenstein:2004ne,Dvali:2000hr,Gregory:2000jc,Damour:2002ws,Blas:2007zz,Carroll:2003wy,Dvali:2008em}. One of them employs spontaneous breaking of Lorentz symmetry by space-time dependent condensates of scalar fields \cite{Rubakov:2004eb,Dubovsky:2004sg} in order to give a mass to gravitons\footnote{See Refs.~\cite{Rubakov:2008nh,Bebronne:2009iy} for reviews.}, which is the reason to call these models \emph{massive gravity} models.

If the gravitons were to have a mass, one would expect the potential of a static source to have a Yukawa shape effectively cutting off the gravitational interaction at distances larger than the inverse graviton mass \cite{Goldhaber:1974wg,Talmadge:1988qz,Smullin:2005iv}. Absence of Lorentz invariance makes these constraints weaker than one would expect in a Lorentz-invariant theory; there is a class of massive gravity models in which the Newtonian potential remains unmodified in the linear approximation \cite{Dubovsky:2005dw}, implying that Solar system constraints are satisfied for rather large graviton mass.

Recently, a class of static vacuum spherically symmetric solutions in massive gravity theories were obtained both analytically and numerically \cite{Bebronne:2009mz}. These solutions depend on two integration constants instead of one in GR: the mass $M$ (or, equivalently, the Schwarzschild radius) and an additional scalar charge $S$. Depending on the parameters of the model considered and the signs of $M$ and $S$, these modified solutions may have both attractive and repulsive (anti-gravitating) behavior at large distances. Those with attractive behavior may mimics the presence of dark matter. Another peculiarity of these solutions is that their asymptotic behavior is different from that obtained in the linear perturbation theory. Despite the obvious interest of these solutions, the underlying hypothesis of spherical symmetry makes their use limited.

The aim of this paper is to understand how these modified solutions emerge from a quasi-linear approximation of massive gravity theories. Beside pure curiosity, the answer to this question sheds lights on the shape of the Newtonian potential produced by any arbitrary mass distribution by identifying the mechanism and equations responsible for these modified solutions. One will then be able to determine, for example, the gravitational potential in a galaxy which is made of two contributions: one contribution coming from the spherically symmetric bulge, and one coming from the axially symmetric galaxy disk. Because of the non-linearity of the equations responsible for these modified solutions, these two contributions cannot be considered separately as in linearized GR. 

By studying the perturbations up to second order about the flat vacuum solution, we found two classes of static spherically symmetric solutions. The first one is a new class of solutions corresponding to solutions of a four-derivative equation for the Newtonian potential. They have a Yukawa shape effectively cutting off the gravitational interaction at distances larger than the inverse graviton mass. The second class of solutions correspond to the modified solutions found in Ref.~\cite{Bebronne:2009mz}.

This paper is organized as follows. In Sect.~\ref{sc:intro-model} we briefly review the massive gravity models and summarize previous results about the Newtonian approximation and the spherically symmetric vacuum solutions. In Sect.~\ref{sc:quasi-lin}, we introduce notations and discuss the quasi-linear equations which will enable us to recover the modified spherically symmetric vacuum solutions. Finally, in Sect.~\ref{sc:static_sol} we discuss the two classes of solutions. For the second class, we show that our quasi-linear static equations reproduce the modified Newtonian potential obtained in Ref.~\cite{Bebronne:2009mz}. Sect.~\ref{sc:sum} contains the summary and discussion of our results.

\section{The massive gravity model}
\label{sc:intro-model}

In this paper, we consider massive gravity models described by the following action \cite{Dubovsky:2004sg},
\begin{eqnarray} \label{eq:action}
\mathcal{S} = \int \dif^4 x \sqrt{- g} \left[ - \Mpl \mathcal{R} + \mathcal{L}_m + \Lambda^4 \mathcal{F} \left( X, W^{ij} \right) \right] .
\end{eqnarray}
The first two terms are the usual Einstein-Hilbert term and the Lagrangian of the minimally coupled ordinary matter; they comprise the standard GR action. The third term describes four scalar fields $\phi^\mu$ whose mixing with the metric give a mass to gravitons. These fields, known as the Goldstone fields, are minimally coupled to gravity through a derivative coupling
\begin{eqnarray*}
X &=& g^{\mu\nu} \partial_\mu \phi^0 \partial_\nu \phi^0 , \\
W^{ij} &=& \left( g^{\mu\nu} - \dfrac{\partial^\mu \phi^0 \partial^\nu \phi^0}{X} \right) \partial_\mu \phi^i \partial_\nu \phi^j .
\end{eqnarray*}
The Goldstone fields $\phi^\mu$ have dimension of length so that $X$ and $W^{ij}$ are dimensionless. The constant $\Lambda$ is a UV cutoff with dimension of mass. Massive gravity models (\ref{eq:action}) are understood as low-energy effective theories valid for energies below $\Lambda$.

The Goldstone fields have space-time dependent vacuum expectation values that break spontaneously the Lorentz symmetry
\begin{eqnarray} \label{eq:vacuum}
g_{\mu\nu} = \eta_{\mu\nu} , & \phi^0 = a t , & \phi^i = b x^i .
\end{eqnarray}
The constants $a$ and $b$ are determined by the requirement that the energy-momentum tensor associated with the four scalar fields vanishes in Minkowski background. Once a function $\mathcal{F}$ is chosen, these constants may be set to one by a redefinition of the fields $\phi^\mu$, which we assume to be the case in what follows, $a = b = 1$.

We consider only functions $\mathcal{F}$ which are invariant under rotations of the Goldstone fields $\phi^i$ in the internal space (i.e., functions depending on $W^{ij}$ through three combinations \mbox{$w_n = \tx{Tr} \left( W^n \right) $ with $n=1,2,3$}). For these functions, the background (\ref{eq:vacuum}) preserves the rotational symmetry. 

Finally, the Goldstone action with functions $\mathcal{F}$ depending only on $X$ and $W^{ij}$ is invariant under the following symmetry
\begin{eqnarray*}
\phi^i \rightarrow  \phi^i + \chi^i \left( \phi^0 \right),
\end{eqnarray*}
where $\chi^i$ are arbitrary functions of $\phi^0$. Because of this symmetry, the perturbations about the vacuum solution (\ref{eq:vacuum}) are non-pathological, i.e. there are neither ghost nor rapid instabilities \cite{Dubovsky:2004sg}. The spectrum consist of two tensor modes (graviton polarizations) only, which are, in general, massive. The graviton mass scale is $m \sim \Lambda^2 / \Mplu$. 

Before discussing quasi-linear static solutions, let us briefly resume the previous results concerning the Newtonian approximation \cite{Dubovsky:2005dw} and the Schwarzschild solutions \cite{Bebronne:2009mz} of theories described by the action (\ref{eq:action}).

\subsubsection*{Newtonian approximation}

For massive gravity models described by (\ref{eq:action}), the gravitational potential of a static source, in the linearized approximation, has the form \cite{Dubovsky:2005dw}
\begin{eqnarray} \label{eq:Newton_lin}
\Phi = M G \left( - \dfrac{1}{r} + \mu^2 r \right) ,
\end{eqnarray}
where $G = \left( 8 \pi \Mpl \right)^{-1}$ is Newton's constant. The first contribution to this potential is the usual GR term. The second contribution is proportional to a constant $\mu$ which is of order of the graviton mass and whose value depends on the particular form of the function $\mathcal{F}$. This contribution implies that the linearized theory breaks at distances $r \geqslant \left( M G \mu^2 \right)^{-1}$.

\subsubsection*{Spherically symmetric solutions}

Recently, the static vacuum spherically symmetric solutions of massive gravity theories described by (\ref{eq:action}) have been obtained both analytically and numerically \cite{Bebronne:2009mz} for different functions $\mathcal{F}$. In order to get some insight into the behavior of these exact solutions, Ref.~\cite{Bebronne:2009mz} studies the exact solution of models described by the following function
\begin{eqnarray} \label{eq:funct_part}
\mathcal{F} \left( X, W^{ij} \right) &=& \dfrac{12}{\lambda} \left( \dfrac{1}{X} + w_1 \right) \\
& & - \left( w_1^3 - 3 w_1 w_2 + 2 w_3 - 6 w_1 - 12 \right) , \nonumber
\end{eqnarray}     
where $\lambda > 0$ is an arbitrary positive constant. For this class of models, the exact static solution of the equations of motion implies that Newton's potential is
\begin{eqnarray} \label{eq:Newton_exact}
\Phi = - \dfrac{G M}{r} - \dfrac{S}{r^\lambda} ,
\end{eqnarray}
where $S$ is an integration constant whose presence reflects the modification of the gravitational interaction as compared to GR. Similar solutions have been previously found in the context of bi-gravity models \cite{Berezhiani:2008nr}. This potential has an asymptotic behavior different from that obtained in the linear perturbation theory. The origin of this apparent contradiction is twofold.

In the linear perturbation theory, all perturbations are assumed to be of the same order: they are formally assigned a small parameter $\epsilon$ to the first power and the Einstein equations are expanded to the linear order in $\epsilon$. The modified solution (\ref{eq:Newton_exact}) is not of this type, even at large distances from the gravitational source. For instance, in the longitudinal gauge, the perturbations of the Goldstone field $\phi^0$ does not decay as fast as the perturbations of the metric; if one assigns a small parameter $\epsilon$ to the perturbation of $g_{00}$, then the perturbation of $\phi^0$ is of order $\sqrt{\epsilon}$ rather than $\epsilon$. Therefore, the exact solution of the static, spherically symmetric, massive gravitational field equations is said to be \emph{non-linear}.

Another difference between the exact solution (\ref{eq:Newton_exact}) and the solution (\ref{eq:Newton_lin}) of the linearized equations is that the former is static, while in the latter only gauge-invariant metric perturbations are static \cite{Dubovsky:2005dw}; the perturbation of the Goldstone field $\phi^0$ only appears in the linearized theory with a time derivative which may be viewed as an accretion of a fluid with zero energy-momentum tensor. Therefore, the two solutions (\ref{eq:Newton_lin}) and (\ref{eq:Newton_exact}) simply describe different physical contexts.

\section{Quasi-linear approximation}
\label{sc:quasi-lin}

Despite the obvious interest of the modified solution (\ref{eq:Newton_exact}), the underlying hypothesis of spherical symmetry makes its use limited; it cannot be used to describe the potential induced by an axially symmetric galaxy. It is therefore important to understand how this modified solution emerges from a quasi-linear theory in order to generalize it to any mass distribution. 

To achieve this, we consider perturbations of the metric and Goldstone fields up to second order about the flat vacuum solution (\ref{eq:vacuum})
\begin{eqnarray*}
g_{\mu\nu} = \eta_{\mu\nu} + \sum_{a=1}^2 h_{(a)\mu\nu} , & & \phi^\mu = x^\mu + \sum_{a=1}^2 \xi^{\mu}_{(a)} ,
\end{eqnarray*}
where $|h_{(a)\mu\nu}| \sim |\xi^{\mu}_{(a)}| \sim \epsilon^{a}$, with $a = 1, 2$.  These perturbations are parameterized as the following \cite{Mukhanov:1990me}
\begin{eqnarray} \label{eq:pert_notation}
h_{(a)00} &=& 2 \varphi_{(a)} , \nonumber \\
h_{(a)0i} &=& S_{(a)i} + \partial_{i} B_{(a)} , \nonumber \\
h_{(a)ij} &=& 2 \psi_{(a)} \delta_{ij} - 2 \partial_{i} \partial_{j} E_{(a)} + H_{(a)ij} \nonumber \\
& &- \partial_i S_{(a)j} - \partial_j S_{(a)i} , \nonumber \\
\xi^{0}_{(a)} &=& \xi^{0}_{(a)} , \nonumber \\
\xi^{i}_{(a)} &=& \xi_{(a)}^i + \partial_{i} \xi_{(a)} .
\end{eqnarray}
The vector perturbations $S_{(a)i}$, $F_{(a)i}$ and $\xi_{(a)}^{i}$ are transverse, while the tensor perturbations $H_{(a)ij}$ are transverse and traceless. In order to simplify the discussion of second order perturbations, let us fix the gauge at the linearized level to be the longitudinal gauge
\begin{eqnarray*}
B_{(1)} = E_{(1)} = F_{(1)i} = 0 .
\end{eqnarray*}
This choice for the linearized approximation will greatly simplify the coming calculations without fixing completely the gauge in the perturbation theory. One is still allowed to perform second order gauge transformations
\begin{eqnarray*}
x^\mu \rightarrow x^\mu - \gamma^\mu &\tx{with}& |\gamma^\mu| \sim \epsilon^2 ,
\end{eqnarray*}
which leave the theory unchanged \cite{Bruni:1996im}. It will then be useful to introduce second order gauge-invariant fields. One vector and two scalar perturbations are gauge degrees of freedom. As a consequence, at the quadratic order there is only two gauge-invariant vector fields
\begin{eqnarray*}
\varpi_{i} = S_{(2)i} + \dot{F}_{(2)i} , & & \sigma_{i} = \xi_{(2)i} - F_{(2)i} ,
\end{eqnarray*}
and four scalar gauge-invariant fields
\begin{eqnarray*}
\begin{array}{ll}
\Phi = \varphi_{(2)} - \dot{B}_{(2)} - \ddot{E}_{(2)} , & \Psi = \psi_{(2)} , \\
\Xi^0 = \xi^0 _{(2)} - B_{(2)} - \dot{E}_{(2)} , & \Xi = \xi_{(2)} -  E_{(2)} .
\end{array}
\end{eqnarray*}
The tensor perturbation $H_{(2)ij}$ is also gauge invariant. In order to increase the readability of the coming calculations, the subscript $(a)$ will be omitted from this point without possible confusion between first and second order perturbations.

Usual matter fields are described by an energy-momentum tensor $\mathcal{T}_{\mu\nu}$. It is convenient to parameterize this tensor in the following way
\begin{eqnarray*}
\mathcal{T}_{\mu\nu} &=& \left( \rho + p \right) v_{\mu} v_{\nu} - g_{\mu\nu} p + \pi_{\mu\nu} + \left( v_{\mu} q_{\nu} + v_{\nu} q_{\mu} \right),
\end{eqnarray*}
where $\rho$ and $p$ are the matter density and pressure measured by a comoving observer of velocity $v_\mu$, $q_\mu$ is the energy flux perpendicular to $v_\mu$ ($v^\mu q_\mu = 0$) and $\pi_{\mu\nu}$ is the anisotropic pressure tensor ($v^\mu \pi_{\mu\nu} = \pi_{\mu}^\mu = 0$). For the vacuum solution (\ref{eq:vacuum}), the energy-momentum tensor is zero and the affine parameter of the observer may be chosen such that $v_\mu = \left( 1, 0, 0, 0 \right)$.

Small perturbations $\delta \mathcal{T}_{\mu\nu}$ of the matter energy - momentum tensor produce metric $\delta g_{\mu\nu} = g_{\mu\nu} - \eta_{\mu\nu}$ and Goldstone $\delta \phi^\mu = \phi^\mu - x^\mu$ perturbations about the vacuum. In other words, the metric and Goldstone perturbations are sourced by $\delta \mathcal{T}_{\mu\nu}$. Since the modified solution (\ref{eq:Newton_exact}) behaves as $| \delta \phi^0 |^2 \sim | \delta g_{00} | \sim M / r$, let us assume that the perturbations of the gravitational source appear at second order only, i.e., $|\delta \rho| \sim |\delta p| \sim |\delta q_\mu| \sim |\delta \pi_{\mu\nu}| \sim \epsilon^2$. As a consequence, $\delta q_0 = \delta \pi_{0\nu} = 0$ and the second order energy - momentum tensor reads
\begin{eqnarray*}
\delta \mathcal{T}_{00} = \delta \rho , & \delta \mathcal{T}_{0i} = \delta q_i , & \delta \mathcal{T}_{ij} = \delta_{ij} \delta p + \delta \pi_{ij} .
\end{eqnarray*}
The energy flux $\delta q_i$ and the anisotropic pressure $\delta \pi_{ij}$ can be parameterized in the following way,
\begin{eqnarray*}
\delta q_i &=& \zeta_i + \partial_i \zeta , \\
\delta \pi_{ij} &=& \left( 3 \partial_i \partial_j - \delta_{ij} \partial_k^2 \right) \pi + \partial_i \pi_j + \partial_j \pi_i + \pi_{ij} ,
\end{eqnarray*}
where the vector perturbations $\zeta_i$ and $\pi_i$ are transverse, while the tensor perturbation $\pi_{ij}$ is transverse and traceless.

With these notations, the energy-momentum conservation reads
\begin{eqnarray*}
\dot{\delta \rho} = \partial_i^2 \zeta , & \dot{\zeta} = \delta p + 2 \partial_i^2 \pi , & \dot{\zeta_i} = \partial_j^2 \pi_i .
\end{eqnarray*}
The ten independent components of the energy-momentum tensor are expressed trough four gauge-invariant scalars $\delta\rho$, $\delta p$, $\zeta$ and $\pi$, four vector degrees of freedom in the form of two transverse gauge-invariant vectors $\zeta_i$ and $\pi_i$, and two tensor degrees of freedom represented by the transverse and traceless gauge-invariant tensor $\pi_{ij}$.

\subsection{First order perturbations}

With the account of the previous notations for the metric and Goldstone perturbations, the linearized Einstein equations split into scalar, vector and tensor equations. These tree sectors could be considered separately\footnote{See the appendix of Ref.~\cite{Bebronne:2007qh} for details about the linearized equations of theories described by (\ref{eq:action}).}.

Without matter fields at the linearized level to source the tensor and vector perturbations, the solutions to the first order equations in these two sectors reads
\begin{eqnarray} \label{eq:lin_sol_1}
S_{i} = \xi^{i} = H_{ij} = 0 .
\end{eqnarray}

The key observation which will enable us to recover the exact, static, potential (\ref{eq:Newton_exact}) from a quasi-linear approximation lies in the scalar sector. Since the perturbation $\xi^{0}$ of the Goldstone field $\phi^0$ only appear at the linearized level with a time derivative, the solution to the linearized scalar equations reads (recall that there is no first order matter source)
\begin{eqnarray} \label{eq:lin_sol_2}
\varphi = \psi = \xi = \dot{\xi}^0 = 0 , 
\end{eqnarray}
implying that $\xi^0$ is an arbitrary function of the space coordinates $\xi^0 = \xi^0 \left( x^i \right)$. This function is not fixed by the first order equations, although one usually assumes from them that $\xi^0 = 0$. Still, it may be that this arbitrary function is determined by higher-order equations. Let us therefore not fix it at this stage, and study the second order approximation.

\subsection{Second order perturbations}

In order to discuss the second order equations, let us assume that the first order solution is given by (\ref{eq:lin_sol_1}) and (\ref{eq:lin_sol_2}) with $\xi^{0} \left( x^i \right)$ to be determined. As for the linearized equations, the second order equations split into scalar, vector and tensor equations. However, each of these three sectors contain contributions proportional to $\xi_0^{2}$.

\subsubsection*{The tensor sector}

The second order tensor perturbations satisfy the following equation
\begin{eqnarray*}
& & \left( \partial_\mu \partial^\mu + m_2^2 \right) H_{ij} + \dfrac{2 \pi_{ij}}{\Mpl}= m_2^2 \left[ \dfrac{\delta_{ij}}{2} \left( \delta_{mn} - \dfrac{\partial_n \partial_m}{\partial_k^2} \right) \right. \\
& & + \delta_{in} \left( \dfrac{\partial_j \partial_m}{\partial_k^2} - \dfrac{\delta_{jm}}{2} \right) + \delta_{jn} \left( \dfrac{\partial_i \partial_m}{\partial_k^2} - \dfrac{\delta_{im}}{2} \right) \\
& & \left. - \dfrac{\partial_i \partial_j}{2 \partial_k^2} \left( \delta_{nm} + \dfrac{\partial_n \partial_m}{\partial_k^2} \right) \right] \partial_{n} \xi_0 \partial_{m} \xi_0 ,
\end{eqnarray*}
where $m_2^2 \propto \Lambda^2 / \Mplu$ is the graviton mass (there are five mass parameters $m_i$ with $i = 0, \ldots, 4$; see Ref.~\cite{Bebronne:2007qh} for a precise expression of these masses in terms of $\mathcal{F} ( X, W^{ij} )$ and its derivatives). This equation describes gravitational waves $H_{ij}$ emitted by the Goldstone perturbation $\xi^0$ and a source with transverse and traceless anisotropic pressure $\pi_{ij}$.

\subsubsection*{The vector sector}

The second order vector perturbations satisfy the two following relations
\begin{eqnarray}
0 &=& \partial_k^2 \varpi_{i} - \dfrac{2 \zeta_i}{\Mpl} , \nonumber \\
0 &=& m_2^2 \left[ \sigma_i + \dfrac{\partial_j}{\partial_k^2} \left( \delta_{in} - \dfrac{\partial_i \partial_n}{\partial_k^2} \right) \partial_{n} \xi^0 \partial_{j} \xi^0 \right] \label{eq:vector} .
\end{eqnarray}
The first of these two equations gives the metric perturbation $\varpi_{i}$ while the second fixes $\sigma_i$ once $\xi^0$ is known.

\subsubsection*{The scalar sector}

The second order scalar sector is described by the following four equations
\begin{eqnarray}
0 &=& 2 \partial_i^2 \Psi - \dfrac{\delta \rho}{\Mpl} + m_0^2 \left[ - \dot{\Xi}^0 + \Phi + \dfrac{1}{2} \left( \partial_i \xi^0 \right)^2 \right] \nonumber \\
& & - m_4^2 \left[ 3 \Psi + \partial_i^2 \Xi + \dfrac{1}{2} \left( \partial_{i} \xi^0 \right)^2 \right] , \label{eq:Scalaire1} \\
0 &=& \partial_i^2 \dot{\Psi} - \dfrac{\dot{\delta \rho}}{2 \Mpl} , \label{eq:Scalaire2} \\
0 &=& 6 \ddot{\Psi} + 2 \partial_i^2 \left( \Phi - \Psi \right) - \dfrac{3 \delta p}{\Mpl} + \left( 3 m_3^2 - m_2^2 \right) \left[ 3 \Psi + \partial_i^2 \Xi \right. \nonumber \\
& & \left. + \dfrac{1}{2} \left( \partial_{i} \xi^0 \right)^2 \right] + 3 m_4^2 \left[ \dot{\Xi}^0 - \Phi - \dfrac{1}{2} \left( \partial_i \xi^0 \right)^2 \right] , \label{eq:Scalaire3} \\
0 &=& \Phi - \Psi + \dfrac{3 \pi}{\Mpl} \nonumber \\
& & + m_2^2 \left[ \Xi + \left( \dfrac{3 \partial_i \partial_j}{\partial_k^4} - \dfrac{\delta_{ij}}{\partial_k^2} \right) \dfrac{\partial_{i} \xi^0 \partial_{j} \xi^0}{4} \right] . \label{eq:Scalaire4}
\end{eqnarray}
This system of four equations describe the four potential $\Phi$, $\Psi$, $\Xi$ and $\dot{\Xi}^0$ emitted by the Goldstone field $\xi^0 = \xi^0 \left( x^i \right)$ and a source with matter and pressure densities $\delta \rho$ and $\delta p$.

At this point, the ten second order Einstein equations have been written and one has to conclude that $\xi^0$ is not determined by second order equations. Moreover, the second order equations do only constraint the value of $\dot{\Xi}^0$, and therefore the value of $\Xi^0$ modulo an arbitrary function of the space coordinates. Consequently, the third-order equations contain ten independent, third-order, gauge-invariant fields\footnote{There are six third-order, gauge-invariant, fields coming from the ``pure'' gravitational sector and four from the Goldstone sector.} plus two arbitrary functions of the space coordinates corresponding to the first and second order initial conditions on $\phi^0$. Hence, these two arbitrary functions wont be constrained by third-order equations either. One may convince oneself that, at each order, the Einstein equations determine $\phi^0$ modulo an arbitrary function which is not fixed by any equation. This arbitrary function may then be chosen to be zero\footnote{The addition of higher-derivative terms in the theory should lift the ambiguity concerning these arbitrary functions.}.

\section{Static quasi-linear solutions}
\label{sc:static_sol}

Although the previous argument seems valid for any arbitrary field configuration, it has a loophole. Indeed, let us search for static solutions to the second order equations, with
\begin{eqnarray*}
\delta \rho = \delta \rho \left( x^i \right) &\tx{and}& \delta p = 0 .
\end{eqnarray*}
Then, equation ($\ref{eq:Scalaire2}$) reduces to an identity and $\Xi^0$ disappears from the equations of motions\footnote{This field only appear in the second order Einstein equation with a time-derivative.}. One is therefore left with three scalar second order equations for $\Psi$, $\Phi$ and $\Xi$, and the second order Einstein equations may be solved by assuming $\xi^0 = 0$. However, by doing this one will miss the branch of solutions which reproduces the exact, modified Newtonian potential (\ref{eq:Newton_exact}).

For static solutions, there is actually a fourth equation which as to be satisfied by second order perturbations. This equation correspond to the longitudinal part of the third-order approximation of the following Einstein equation
\begin{eqnarray*}
\mathcal{G}_{0i} = \dfrac{1}{\Mpl} \left( \mathcal{T}_{0i} + t_{0i} \right), 
\end{eqnarray*}
where $\mathcal{G}_{\mu\nu}$ is the Einstein tensor and $t_{\mu\nu}$ is the energy-momentum tensor of the four Goldstone fields. Indeed, for a static solution, this third-order equation reduces to
\begin{eqnarray} \label{eq:third_order}
0 &=& \partial_i \left\{ \partial_i \xi^0 \left[ \dfrac{1}{2} \left( m_2^2 - m_0^2 - m_3^2 + 2 m_4^2 \right) \left( \partial_k \xi^0 \right)^2 \right. \right. \nonumber \\
& & \left. + m_2^2 \Psi + \left( m_4^2 - m_0^2 \right) \Phi + \left( m_4^2 - m_3^2 \right) \left( 3 \Psi + \partial_k^2 \Xi \right) \right] \nonumber \\
& & \left. + m_2^2 \partial_k \xi^0 \left[ \partial_k \partial_i \Xi + \frac{1}{2} \left( \partial_k \sigma_{i} + \partial_i \sigma_{k} \right) \right] \right\} .
\end{eqnarray}
Since there is no third-order fields in this equation\footnote{The third-order fields enter this equation with a time-derivative and therefore disappear when considering static solutions.}, any solution to the second order equations has to satisfy this third-order equation as well. Moreover, this equation implies that there are two branches of static solutions, as already stressed in Ref.~\cite{Bebronne:2009mz}. The first branch corresponds to solutions with $\xi^0 = 0$ for which equation (\ref{eq:third_order}) reduces to an identity. The second branch, characterized by $\xi^0 \neq 0$, is the one studied in Ref.~\cite{Bebronne:2009mz}. Let us discuss these two branches separately.

\subsection[First class of solutions]{Solutions with $\xi^0 = 0$}

If one considers solutions for which $\xi^0 = 0$, then the second order scalar equations reduce to a system of three equations (\ref{eq:Scalaire1}), (\ref{eq:Scalaire3}) and (\ref{eq:Scalaire4}) for $\Psi$, $\Phi$ and $\Xi$. These equations imply that Newton's potential satisfies a four-derivative equation
\begin{eqnarray*}
0 &=& \left( 3 m_4^2 - m_0^2 - 2 \partial_i^2 \right) \Phi + \dfrac{\delta \rho}{\Mpl} \\
& & + \dfrac{\left( m_2^2 + m_4^2 - 3 m_3^2 \right) \left[ 3 m_4^2 m_2^2 + \left( m_4^2 - 2 m_2^2 \right) \partial_i^2 \right]}{\left[ m_2^2 \left( 3 m_3^2 - m_2^2 \right) + \left( m_3^2 - m_2^2 \right) \partial_i^2 \right]} \Phi , 
\end{eqnarray*}
whose solution for a point-like source $\delta \rho = M \delta^3 \left( x \right)$ may be approximated at small and large distances by
\begin{eqnarray*}
\Phi \simeq - \dfrac{M G}{r} , &\tx{for}& r \ll m^{-1} , \\
\Phi \simeq - \eta \dfrac{M G}{r} \tx{e}^{- | m_2^2 \tau| r} , &\tx{for}& r \gg m^{-1} ,
\end{eqnarray*}
where $\eta$ and $\tau$ are dimensionless combinations of the masses $m_i^2$. One therefore concludes that the Newtonian potential deduced from a static solutions with $\xi^0 = 0$ is given by its value in GR at small distances and is cut off at distances larger than $m^{-1}$.

\subsection[Second class of solutions]{Solutions with $\xi^0 \neq 0$}

The solutions with $\xi^0 \neq 0$ are those which reproduce the modified potential (\ref{eq:Newton_exact}). Because of the complexity of the third-order equation (\ref{eq:third_order}), let us concentrate on spherically symmetric solutions. 

For static, spherically symmetric solutions with $\xi^0 \neq 0$, equation (\ref{eq:third_order}) reduces to
\begin{eqnarray}
0 &=& \dfrac{1}{2} \left( m_2^2 - m_0^2 - m_3^2 + 2 m_4^2 \right) \left( \partial_r \xi^0 \right)^2 + \left( m_4^2 - m_0^2 \right) \Phi  \nonumber \\
& & + \left( m_4^2 - m_3^2 \right) \left( 3 \Psi + \partial_r^2 \Xi + \dfrac{2}{r} \partial_r \Xi \right) \nonumber \\
& & + m_2^2 \left( \Psi + \partial_r^2 \Xi \right) \label{eq:Spheric_1}
\end{eqnarray}
while the three equations (\ref{eq:Scalaire1}), (\ref{eq:Scalaire3}) and (\ref{eq:Scalaire4}) read 
\begin{eqnarray}
0 &=& 2 \partial_r^2 \Psi + \dfrac{4}{r} \partial_r \Psi - \dfrac{\delta \rho}{\Mpl} + m_0^2 \left[ \Phi + \dfrac{1}{2} \left( \partial_r \xi^0 \right)^2 \right] \nonumber \\
& & - m_4^2 \left[ 3 \Psi + \partial_r^2 \Xi + \dfrac{2}{r} \partial_r \Xi + \dfrac{1}{2} \left( \partial_{r} \xi^0 \right)^2 \right] , \label{eq:Spheric_2} \\
0 &=& 2 \left( \partial_r^2 + \dfrac{2}{r} \partial_r \right) \left( \Phi - \Psi \right) + \left( 3 m_3^2 - m_2^2 \right) \left[ 3 \Psi + \partial_r^2 \Xi \right. \nonumber \\
& & \left. + \dfrac{2}{r} \partial_r \Xi + \dfrac{1}{2} \left( \partial_{r} \xi^0 \right)^2 \right] - 3 m_4^2 \left[ \Phi + \dfrac{1}{2} \left( \partial_r \xi^0 \right)^2 \right] , \label{eq:Spheric_3} \\
0 &=& \left( \partial_r^2 - \dfrac{1}{r} \partial_r \right) \left( \Phi - \Psi + m_2^2 \Xi \right) + \dfrac{m_2^2}{2} \left( \partial_r \xi^0 \right)^2 . \label{eq:Spheric_4}
\end{eqnarray}
This is a system of four equations for $\Phi$, $\Psi$, $\Xi$ and $\left( \partial_r \xi_0 \right)^2$. It can be solved as follows. Adding equations (\ref{eq:Spheric_2}), (\ref{eq:Spheric_3}) and (\ref{eq:Spheric_4}) to equation (\ref{eq:Spheric_1}) gives a closed equation for $\Phi + r \partial_r \Psi$
\begin{eqnarray} \label{eq:PhiPsi}
0 = \dfrac{1}{r} \partial_r \left( \Phi + r \partial_r \Psi \right) - \dfrac{\delta \rho}{2 \Mpl},
\end{eqnarray}
whose solution reads
\begin{eqnarray}
\Phi + r \partial_r \Psi &=& \omega \left( r \right) , \label{eq:Psi-Phi} \\
\omega \left( r \right) &\equiv& 4 \pi G \int_{0}^{+\infty} \dif r^{\prime} \int \dif r \,\, r \delta \left( r - r^\prime \right) \delta \rho \left( r^\prime \right) . \nonumber
\end{eqnarray}
For a point-like source $\delta \rho = M \delta^3 \left( x \right)$ and $\omega \left( r \right) = - G M \delta \left( r \right)$, where $\delta \left( r \right)$ is the delta function in spherical coordinates\footnote{The one dimensional delta function $\delta \left( r \right)$ is related to the three dimensional one through the usual relation \begin{eqnarray*}
\int_{-\infty}^{+\infty} \delta^3 \left( x \right) \dif^3 x = 1 = \int_0^{+\infty} \delta \left( r \right) \dif r &\Longrightarrow&
\delta^3 \left( x \right) = \dfrac{1}{4 \pi r^2} \delta \left( r \right) .\end{eqnarray*}}. It is worth noting that equation (\ref{eq:PhiPsi}) is the linearized version of the relation $\partial_r ( g_{00} \, g_{rr} ) = 0$ characterizing the Schwarzschild solution of GR in the usual spherical coordinates.

Solving the three remaining equations for arbitrary functions $\mathcal{F}$ of $X$ and $W^{ij}$ is a difficult task which will not be achieved here. However, let us discuss the solution of these three equations for two classes of models.

\subsubsection*{First class of models}

The first class of models which will be addressed here correspond to models described by the function (\ref{eq:funct_part}) discussed in Ref.~\cite{Bebronne:2009mz}. We should stress here that this class of models is only particular because we know how to solve analytically the exact, static, spherically symmetric equations in these models.

At the quasi-linear level, the static equations are easily solved because of the following relations between the mass parameters 
\begin{eqnarray*}
m_0^2 = 3 m_4^2 , & m_2^2 = m_4^2 \left( 2 + \lambda \right) , & m_3^2 = m_4^2 \left( 1 + \lambda \right).
\end{eqnarray*}
Indeed, having the solution of equation (\ref{eq:PhiPsi}) in mind, equations (\ref{eq:Spheric_1}), (\ref{eq:Spheric_2}) and (\ref{eq:Spheric_3}) imply that
\begin{eqnarray*}
\left( \partial_r - \dfrac{\lambda}{r} \right) \left( \partial_r \Xi + r \Psi \right) = \omega \left( r \right) .
\end{eqnarray*}
The solution to this equation reads
\begin{eqnarray}
\partial_r \Xi + r \Psi &=& \kappa \left( r \right) , \label{eq:Xi-Psi} \\
\kappa \left( r \right) &\equiv& \int_{0}^{+\infty} \dif r^{\prime} \Theta \left( r - r^\prime \right) \left( \dfrac{r}{r^\prime} \right)^\lambda \omega \left( r^\prime \right) . \nonumber
\end{eqnarray}
For a point-like source, $\kappa \left( r \right) = 0$. The remaining two equations imply then a closed inhomogeneous equation for the Newtonian potential $\Phi$
\begin{eqnarray*}
0 &=& \left( \partial_r^2 + \dfrac{2 + \lambda}{r} \partial_r + \dfrac{\lambda}{r^2} \right) \Phi - \dfrac{\delta \rho \left( r \right)}{2 \Mpl} - \dfrac{\lambda}{r^2} \omega \left( r \right) \\
& & + \dfrac{3 \lambda m_4^2}{2} \dfrac{2 + \lambda}{r} \kappa \left( r \right).
\end{eqnarray*}
For a point-like source, the solution of this equation is exactly the modified potential (\ref{eq:Newton_exact})
\begin{eqnarray} \label{eq:pot_func}
\Phi = - \dfrac{G M}{r} - \dfrac{r_s}{r} - \dfrac{S}{r^{\lambda}}, 
\end{eqnarray}
where $r_s$ and $S$ are two integration constants. Making use of this solution and integrating equations (\ref{eq:Psi-Phi}), (\ref{eq:Xi-Psi}) and (\ref{eq:Spheric_4}) one find the three other fields
\begin{eqnarray*}
\Psi &=& G M \left( \delta \left( r \right) - \dfrac{1}{r} \right) - \dfrac{r_s}{r} - \dfrac{S}{\lambda r^{\lambda}} , \\
\Xi &=& \left( G M + r_s \right) r - \dfrac{S}{\lambda \left( \lambda - 2 \right) r^{\lambda-2}} , \\
\left( \partial_r \xi^0 \right)^2 &=& - \dfrac{2}{m_2^2} \left[ \dfrac{S}{r^{\lambda+2}} \left[ 2 - \lambda \left( \lambda + 1 \right) \right] - \dfrac{3 G M \delta \left( r \right)}{r^2} \right] \\
& & - 2 G M \delta \left( r \right) + \dfrac{2 G M}{r} + \dfrac{2 r_s}{r} + \dfrac{2 S}{r^{\lambda}} .
\end{eqnarray*}

\subsubsection*{Second class of models}

The second class of models which will be addressed here correspond to attractors of the cosmological evolution \cite{Dubovsky:2005dw}. Their action depends on the variables $X$ and $W^{ij}$ through a single variable $Z^{ij} \equiv X^\gamma W^{ij}$
\begin{eqnarray} \label{eq:funcZ}
\mathcal{F} = \mathcal{F} \left( Z^{ij} \right), 
\end{eqnarray}
where the constant $\gamma$ is a free parameter. These models have been studied intensively \cite{Dubovsky:2005dw,Bebronne:2007qh,Bebronne:2008tr}. In particular, it has been shown that for $- 1 < \gamma < 0$ and for $\gamma = 1$ the cosmological perturbations in these models behave identically to those in GR \cite{Bebronne:2007qh}. For other values of $\gamma$ the behavior of the cosmological perturbations may or may not reproduce that of GR depending on the initial conditions.

Another reason to study models characterized by a function (\ref{eq:funcZ}) comes from the analysis of the linearized approximation. For these models, the $\mu$ parameter in equation (\ref{eq:Newton_lin}) is zero \cite{Dubovsky:2005dw} with consequences that Newton's potential remains unmodified at the linearized level.

Still, at the non-linear level, the spherically symmetric vacuum solutions have similar behavior \cite{Bebronne:2009mz} as those described by the modified potential (\ref{eq:Newton_exact}). Indeed, equations (\ref{eq:Spheric_1}), (\ref{eq:Spheric_2}) and (\ref{eq:Spheric_3}) are easily solved because of the following relations between the mass parameters 
\begin{eqnarray*}
m_0^2 = 3 \gamma m_4^2 , & & m_4^2 = \gamma \left( 3 m_3^2 - m_2^2 \right) ,
\end{eqnarray*}
characterizing these models. Having the solution of equation (\ref{eq:PhiPsi}) in mind, equations (\ref{eq:Spheric_1}), (\ref{eq:Spheric_2}) and (\ref{eq:Spheric_3}) imply that the Newtonian potential satisfies
\begin{eqnarray} \label{eq:pot_F(Z)}
0 = \left( \partial_r^2 + \dfrac{2 + \lambda}{r} \partial_r + \dfrac{\lambda}{r^2} \right) \Phi - \dfrac{\delta \rho \left( r \right)}{2 \Mpl} - \dfrac{\lambda}{r^2} \omega \left( r \right) ,
\end{eqnarray}
Here, $\lambda = 1 - 1 / \gamma$. For a point-like source, the solution of this equation is also given by the modified potential (\ref{eq:pot_func}). The behavior of this solution at $r \rightarrow \infty$ depends on the constant $\gamma$. For $\gamma < 0$ and for $\gamma > 1$, this solution describes asymptotically flat space-time.

\section{Discussion}
\label{sc:sum}

The aim of this paper was to understand how the modified Newtonian potential (\ref{eq:Newton_exact}), inferred from static spherically symmetric solutions of massive gravity theories, arises from a quasi-linear perturbation theory. By studying the Einstein equations up to second order perturbations, one has identified two branches of static solutions. In the first-order longitudinal gauge, the key difference between these two branches is the perturbation $\xi^0$ of the Goldstone scalar field $\phi^0$.

The first branch of solutions is a new one corresponding to solutions with $\xi^0 = 0$. They are solutions of a four-derivative equation for the Newtonian potential. They reproduce the potential of GR at small distances with a Yukawa shape effectively cutting off the gravitational interaction at distances larger than $m^{-1}$.

The second branch of solutions, characterized by $\xi^0 \neq 0$, are those who reproduce the modified potential (\ref{eq:Newton_exact}). The quasi-linear equations have been solved with $\xi^0 \neq 0$ in two different classes of models. The first one corresponds to models characterized by the function (\ref{eq:funct_part}). The static vacuum spherically symmetric solutions were already known in these models. The second class of models correspond to attractor of the cosmological evolution for which one only has numerical static spherically symmetric solutions. By solving the quasi-linear equations, the present work sheds lights on the static spherically symmetric solutions in this second class of models.

In this second branch of solutions, the Newtonian potential (\ref{eq:pot_func}) depends on two integration constant: the Schwarzschild radius $r_s$ and an additional scalar charge $S$. For usual source, these two integration constants have to be zero. Indeed, for a sphere of constant density $\varrho$ and radius $R$, the energy density is given by $\delta \rho = \varrho \Theta \left( R - r \right)$. For such source, one may show that the solution to equation (\ref{eq:pot_F(Z)}) reads
\begin{eqnarray*}
\Phi &=& \dfrac{3 G M}{2 R} - \dfrac{G M}{r} \Theta \left( r - R \right) - \dfrac{r_s}{r} - \dfrac{S}{r^{\lambda}}  \\
& & + \dfrac{G M}{2 R} \left( \dfrac{r^2}{R^2} - 3 \right) \Theta \left( R - r \right),
\end{eqnarray*}
with $M = 4 \pi \varrho R^3 / 3$. The first term in this equation is a new contribution which may be re-absorbed by a redefinition of time. Since there is no singularity in the center of the source, the only physical solution corresponds to $r_s = S = 0$. Therefore, the scalar charge of an
ordinary source is zero.

It remains an open question how objects with $S \neq 0$ can be created. However, the argument given above does not apply to black holes, especially to super-massive black holes in the centers of galaxies, which may be of primordial origin \cite{Carr:2005bd}. The previous argument does not apply to time-dependent configurations either, so it is possible that a non-zero scalar charge may be acquired during the gravitational collapse.

To conclude, let us stress that the equations presented in this paper may be used to determine the static gravitational potential produced by any arbitrary mass distribution. Although we have mainly discussed spherically symmetric solutions, the potential produced by a galaxy, for example, may be found by solving numerically equations (\ref{eq:vector}), (\ref{eq:Scalaire1}), (\ref{eq:Scalaire3}), (\ref{eq:Scalaire4}) and (\ref{eq:third_order}) altogether with the hypothesis of axial symmetry.

\subsubsection*{Acknowledgments}

I am grateful to Peter Tinyakov for stimulating discussions and comments on this manuscript.

\end{document}